\documentclass[aps,11pt,notitlepage,oneside,onecolumn,nobibnotes,nofootinbib,superscriptaddress,centertags]{revtex4}               

\usepackage{graphicx}

\usepackage{epsfig}
\usepackage{eucal} 
\usepackage{amssymb}
\usepackage{amsfonts}
\usepackage{amsbsy}

\begin{document}


\title{\Large{Internally Electrodynamic Particle Model: Its Experimental Basis} \\ \Large{and Its Predictions}
}
\author{J.X. Zheng-Johansson}
\address{Institute of Fundamental Physics Research, Nyk\"oping, Sweden}

\def\SecIEDmodel{2}
\def\secsolu{3}
\def\Secexp{4}

\def\ip{{i'}}
\def\empty{{\mbox{\tiny${\emptyset}$}}}
\def\min{\mbox{-}}
\def\rb{{\bf{r}}}
\def\v{{\rm{v}}}
\def\vac{{\rm{vac}}}
\def\Ai{A$_{0}$}
\def\Bi{B$_{1}$}
\def\Bii{B$_{2}$}
\def\EMsup{{          {}_{\mbox{\tiny${em}$}}             }}
\def\EM{electromagnetic}
\def\lsub{\mbox{\tiny${l}$}}
\def\alphab{{\pmb{\alpha}}}
\def\sigmab{{\pmb{\sigma}}}
\def\Xim{{\cal X}}
\def\PsimR{\widetilde\Psim}
\def\Psima{\Psim}
\def\el{\mbox{${e}$}}
\def\elsub{\mbox{\tiny${e}$}}
\def\elmsub{\mbox{\tiny{${e}^{\minus}$}}}
\def\elpsub{\mbox{\tiny{${e}^{\p}$}}}

\def\gsub{\mbox{\tiny${\gamma}$}}

\def\Omegavel{\mathbin{{\mit\Omega}\mkern-13.mu^{_{\mbox{$-$}}}\hspace{-0.08cm}{}_d }}
\def\omegavel{\mathbin{{\mit\omega}\mkern-13.mu^{_{\mbox{$-$}}}\hspace{-0.08cm}{}_d }}
\def\Wvel{\Omegavel}
\def\wvel{\omegavel}
\def\nuvel{\mathbin{{\mit\nu}\mkern-13.mu^{_{\mbox{$-$}}}\hspace{-0.08cm}{}_d }}
\def\Nuvel{\mathbin{{\mit\Nu}\mkern-13.mu^{_{\mbox{$-$}}}\hspace{-0.08cm}{}_d }}

\def\Ci{1}
\def\betamt{{\bf{b}}}
\def\kb{{\bar{k}}}
\def\kbf{{\bf{k}}}
\def\Kb{{\bf{K}}}
\def\cb{{\bf{c}}}

\def\pb{{\bar{p}}}
\def\pbf{{\bf{p}}}
\def\Acal{{\cal{A}}}
\def\Bcal{{\cal{B}}}
\def\Ccal{{\cal{C}}}
\def\Vp{V}
\def\m{{{}_{\mbox{-}}}}
\def\Ccal{{\cal{C}}}
\def\p{{{}_{+\hspace{-0.1cm}}}}

\def\psipi{\psi_{\p}(1)}
\def\psipii{\psi_{\p}(2)}
\def\psimi{\psi_{\m}(1)}
\def\psimii{\psi_{\m}(2)}

\def\ai{\alpha(1)}
\def\aii{\alpha^{'}(2)}
\def\bi{\beta^{'}(1)}
\def\bii{\beta(2)}

\def\fa{f_r}
\def\fb{f_\ell}

\def\Ca{C_a}
\def\Cb{C_b}
\def\fbf{{\bf{f}}}
\def\Ocal{{\cal{O}}}
\def\psib{{\pmb{\psi}}}
\def\alphab{{\pmb{\alpha}}}
\def\sigmab{{\pmb{\sigma}}}

\def\Eb{{\bf E}}
\def\Bb{{\bf B}}
\def\ke{\kappa}
\def\nabb{{\pmb{\nabla}}}
\def\nablab{{\pmb{\nabla}}}
\def\vir{{\rm vir}}
\def\psitot{\psi}
\def\jb{{\bf{j}}}
\def\vel{v}
\def\velb{{\bf{v}}}
\def\velsub{\mbox{\tiny{$\vel$}}}

\def\Imtr{I}
\def\citeUnif{1}

\def\App{}
\def\Qcal{{\mathcal{Q}}}
\def\Tcal{{\mathcal{T}}}
\def\Cross{Q}

\def\vphilim{f}
\def\ft{{\mathcal{B}}}
\def\vphibar{\mathbin{\varphi\mkern-12.5mu-}}

\def\vphi{\varphi}
\def\med{{\med}}

\def\Mcal{{\mathfrak{M}}}

\def\Sb{{\bf{S}}}
         \def\xia{{\mathcal{A}}}
\def\tha{\theta}

\def\nb{\bf{n}}
\def\zb{{\bf{z}}^0}
\def\phiv{\varphi}
\def\Lb{{\bf{L}}}
\def\velsub{_{\vel}}

\def\nablab{{\pmb{\nabla}}}
\def\velb{{\pmb{\vel}}}
\def\minus{\mbox{-}}
\def\m{\mbox{-}}

\def\Ab{{\bf{A}}_a}
\def\vel{\upsilon}
\def\Thm{\vartheta}
\def\lb{{\bf l}}
\def\vb{{\bf{v}}}

\def\Rb{{\bf R}}
\def\pd{\partial}
\def\vphi{\varphi}

\def\psiR{\widetilde{\psi}}
\def\psiL{\widetilde{\psi}^{{\rm vir}}}
\def\PhimR{\widetilde{ {\mit \Phi}}}
\def\PsimR{\widetilde{ {\mit \Psi}}}
\def\PsimL{{\widetilde{ {\mit \Psi}}}^{{\rm vir}}}
\def\a{\alpha}
\def\uav{\bar{u}}
\def\D{\Delta}
\def\th{\theta}
\def\r{{\mbox{\tiny${R}$}}}
\def\re{{\mbox{\tiny${R}$}}}
\def\Fmed{F_{{\rm a.med}}}
\def\med{{\rm med}}
\def\Lw{L_{\varphi}}
\def\Fb{{\bf{F}}}

\def\Efb{{\bf{E}}}
\def\Bfb{{\bf{B}}}
\def\Ac{ \varphi}
\def\Xsub{{\mbox{\tiny${X}$}}}
\def\Ysub{{\mbox{\tiny${Y}$}}}
\def\Zsub{{\mbox{\tiny${Z}$}}}

\def\Ksub{{\mbox{\tiny${K}$}}}
\def\W{{\mit \Omega}}
\def\Wd{\W_d{}}
\def\Nu{{\cal V}}
\def\Nud{\Nu_d{}}
\def\Eng{{\cal E}}
\def\eng{{\varepsilon}}
\def\Acuni{\Ac_{{\Ksub}^\dagsup}^{\dagsup}}
\def\unduni{\Ac_{{\Ksub}^\dagger}^{\dagsup}}
\def\Acauni{\Ac_{{\Ksub}^\ddagsup}^{\ddagsup}}
\def\Acunim{{\Ac_{{\Ksub}^\dagsup}^{\dagsup *}}}
\def\undunim{{\Ac_{{\Ksub}^\dagsup}^{\dagsup}}^*}
\def\Acaunim{{\Ac_{{\Ksub}^\ddagsup}^{\ddagsup *}}}
\def\pd{\partial}
\def\Ad{ {\mit \psi}}
\def\psim{ {\mit \psi}}
\def\Kd{K_d{}}
\def\Lam{{\mit \Lambda}}
\def\lam{\lambda}
\def\dagsup{{\mbox{\tiny${\dagger}$}}}
\def\ddagsup{{\mbox{\tiny${\ddagger}$}}}
\def\psimKdK{\psim_{\Ksub,\Kdsub}}
\def\w{\omega{}}
\def\wdlow{\omega_d }
\def\g{\gamma{}}
\def\Phim{{\mathcal C}}
\def\Psim{{\mit \Psi}}
\def\arm{{\rm a}}
\def\brm{{\rm b}}
\def\crm{{\rm c}}
\def\drm{{\rm d}}
\def\erm{{\rm e}}
\def\frm{{\rm f}}
\def\grm{{\rm g}}
\def\hrm{{\rm h}}
\def\lf{\left}
\def\rt{\right}
\def\Kdsub{{\mbox{\tiny${K_d}$}}}
\def\psimkd{\psim_{\kdsub}}
\def\psimKd{\psim_{\Kdsub}}
\def\hquad{ \ \ }
\def\Taum{{\mit \Gamma}}

\begin{abstract}
The internally electrodynamic (IED) particle model was derived based on overall experimental observations,  with  the  IED process  itself being built directly on three experimental facts that, a) electric charges present with  all material  particles,  b) an accelerated  charge generates  electromagnetic waves  according to  Maxwell's equations and the Planck energy equation and  c) source motion produces Doppler effect. 
A set of well-known basic particle equations and properties become predictable based on first principles solutions for the IED process;
several key solutions  achieved are outlined,  including the de Broglie phase wave, de Broglie relations, Schr\"odinger equation, mass, Einstein mass-energy relation, Newton's law of gravity,  single particle self interference, and electromagnetic radiation and absorption; these equations and properties  have  long been broadly experimentally validated or demonstrated. A conditioned solution also predicts  the Doebner-Goldin equation which emerges to  represent a  form of long-sought quantum wave equation including  gravity.
 A critical  review of the key experiments is given which suggests that the IED process underlies the basic particle equations and properties not just sufficiently but also necessarily. (Appeared in: Physics of Atomic Nuclei, 2010, Vol 73, No 3, pp.571-581.)

\end{abstract}
 \maketitle

\section*{\large{1. Open questions. The need for a comprehensive particle model}} 
 
It is  well established today that all material particles exhibit a dual wave and particle property,  hence described as matter waves,   and their motions at quantum scale are governed by the Schr\"odinger, or alternatively the Heisenberg, and the Dirac equations  in the respective velocity regimes.
We are also  faced today with
a range of open questions regarding  particles in the realm of  fundamental physics.     What is waving with the  matter wave ($\psi$) which also dually manifests as a particle?
How does a particle  interfere with itself say  in a double-slit?
How is an  electromagnetic wave   on absorption converted to a portion or the whole of  the mass or energy of a particle, and conversely on emission?
What  is the origin of mass?
Why  do masses attract one another? How does the gravity enter quantum wave equation?   And so forth.

  The ultimate answers to the questions inevitably are intimately interconnected. For example, one can not have answered what is waving without answering at the same time how mass enters in what is waving.
The "wave", "relativistic mass", etc.
are all quantities in a dynamic domain where a basic rule for all is therefore the consistency in energy, or energy conservation.
From an {\it a priori} energy consideration we recognize that the matter wave $\psi $ must represent  an internal degree of freedom  of a particle. In other terms, the waving of $\psi$  can not be  the waving of the mass ($m$) of the particle;  if it were, the particle would have an excess mechanical energy  $m{\dot{\psi}}^2$ which we know it has not. To answer the various questions we inevitably need  a more comprehensive particle model than, approximately speaking,  a statistical point particle picture with a given-for-granted mass.

\section*{2. The IED particle model: The direct experimental basis} 

\def\SecIEDmodel{2}

Based on overall  experimental observations as input information the author recently developed an internally electrodynamic (IED) particle model \cite{Unif1,Unif1Schr,Unif1dBw,Unif1mass,Unif1Radiation,Unif1Dirac,Unif1D-GEqn,Unif1Vacdiel,Unif1Grv,Unif1Vac}  (earlier termed a basic particle formation scheme) which briefly states:
{\it
A single-charged material particle, like the electron, proton,  etc., is constituted of (i) an oscillatory point charge $q$  of a characteristic frequency  $\W$ and   zero rest-mass,  and (ii) the electromagnetic waves  generated by the charge and propagated between the charge and reflecting boundaries }(Fig.1a).
{\it The waves will be subject to a Doppler effect  if the oscillatory charge as a whole, the source, is in motion;
$q$ is an electric charge  in the usual electromagnetic sense and thus obeys the basic laws of electrodynamics; the total  energy of the oscillatory charge  or equivalently  of the electromagnetic waves  is associated with a  (dynamical)  inertial mass obeying the usual laws of mechanics.}
When going down to a deeper level so as to address the mechanical basis for charge oscillation, the origin of mass, etc.,  
the  vacuum is represented as a substantial vacuuonic medium, and  the charge moving in it will be resisted by a medium force  to identify with the usual  inertial force.
\begin{figure} \includegraphics[width=0.55\textwidth]{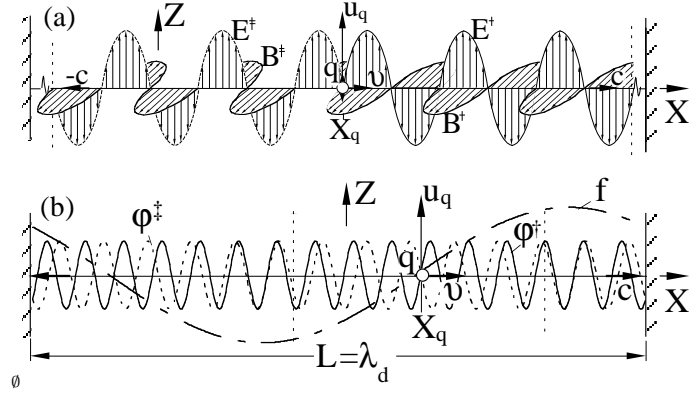}
\vspace{-0.5cm}
\caption{ An  IED particle consisting of an  oscillatory charge  $q(\w)$, travelling at velocity $\vel$ in $+X$-direction,  and the resulting  electromagnetic waves, of electric and magnetic  fields $E^j,B^j$ as in (a),   propagated in $+X$-  ($j=\dagger$) and $-X$-  ($j= \ddagger$)  directions between heavy walls spaced at $L$. In  (b), the $E^j$ fields are plotted as dimensionless functions $\vphi^j$'s given by exact solutions (Sec.  3.1); $f$ is the modulation envelop  of de Broglie phase wave $\psi$ given in  (\ref{eq-dBw1aa})$'$.
}\label{fig1} \end{figure}

The IED process  is itself  built on the experimental facts  embedded in the  set of established basic  laws:
(a) all material particles consist of
 finite amounts of electric charges with accordingly defined spins  \cite{ParticleDataGroup2008};
(b)
an accelerated charge generates electromagnetic waves of electric and magnetic fields $\Eb^j$ and $\Bb^j$ ($j=\dagger,\ddagger$) according to  Maxwell's equations  (JC Maxwell, 1873; H Hertz 1888), 
$$ \displaylines{\refstepcounter{equation}\label{eq-maxwel1}
 \nablab \cdot \Eb^j =\frac{\rho_q^j}{\epsilon},\quad
 \nablab \cdot \Bb^j=0,
\quad
\nablab \times \Bb^j=\mu \jb_q^j  +\frac{1}{c^2}\frac{\pd \Eb^j}{\pd t},
\quad
\nablab \times \Eb^j =- \frac{\pd \Bb^j}{\pd t},  \hfill (\ref{eq-maxwel1})
}$$
  $c$ being  the velocity of light,
with the wave's energy amplitude $\propto \epsilon |\Eb|^2$ ($E=\sqrt{E^{\dagsup}E^{\ddagsup}}$ here) being in nature quantized according to the Planck  equation
\cite{Planck:1900}
$$ \displaylines{\refstepcounter{equation}\label{eq-Plancke}
\eng=n_{ph}\hbar \w, \quad  \ n_{ph}=1,2, \ldots, \hfill (\ref{eq-Plancke})
}$$
(c) the  motion  of source  (the oscillatory charge), at velocity $\vel$, yields
  a Doppler effect (C Doppler, 1842;  \cite{Stark1905,DuMond1949}), with the
 wavevectors and frequencies of the waves generated in the directions 
 parallel  ($j=\dagger$)  and   antiparallel ($j=\ddagger$)  with  $\vel$ displaced from their monochromatic values $K$ and $\W=cK$  to:
$$ \displaylines{\refstepcounter{equation}\label{eq-Doppler}
 k^{j}=\g^{j}K, \quad \w^{j}=ck^j=\g^{j}\W, \quad
\g^{\dagsup}=1/(1-\vel/c), \quad
\g^{\ddagsup}=1/(1+\vel/c),
\hfill (\ref{eq-Doppler})
}$$
and (d) the Newton's laws of motion and (e) the Lorentz force law in respect to the dynamics of the point charge.   (a)-(e) make up the  first principles laws here.

Clearly, the finite charge $q$  in the IED model is a direct mapping of  law (a).
The zero rest mass of $q$, being specific with the IED model, ensures  that  the mass $m$ of the resulting particle correctly is the dynamical consequence of the IED process and is not endowed twice; this is on  equal footing with the well appreciated notion of a zero rest mass of the  electromagnetic waves.
Laws  (b) and (c) are experimentally demonstrated for  electromagnetic waves   emitted  "permanently"  from their sources (charged particles),
hence appearing "external" to the  particles. Yet the same laws (b)-(c) ought naturally to  apply  to the electromagnetic waves internal of the IED particle since they are emitted by the same charge and propagated in the same vacuum; they appear "internal" only in the way that they are repeatedly   re-absorbed by the charge and then re-emitted.
Similarly,  laws (d)-(e) ought to apply to the charge internal of an IED particle as in practice we commonly apply  to other  internal charges, like the charges of an atomic electron and of an atomic nucleus.

\begin{figure}[thtbp] \includegraphics[width=0.3\textwidth]{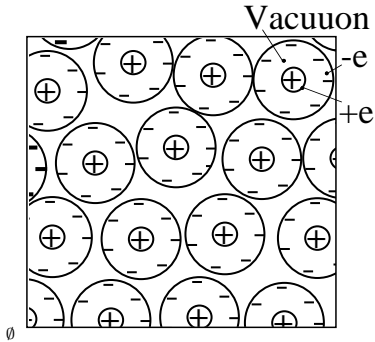}
\vspace{-0.5cm}
\caption{Schematic structure of  vacuum. }\label{fig2} \end{figure}

It suffices to represent  the IED wave process \cite{Unif1,Unif1Schr,Unif1dBw,Unif1mass,Unif1Radiation,Unif1Dirac} with the usual electromagnetic fields governed by laws (b)-(c). Although, a physical construction  of the  vacuum is compelling for addressing issues like the origin of mass (e.g. in \cite{Unif1,Unif1Schr,Unif1mass}),  the mechanical basis of  charge and medium oscillations (e.g. in \cite{Unif1,Unif1Schr,Unif1dBw,Unif1mass,Unif1Dirac,Unif1D-GEqn})
in contrast to an {\it ad hoc} imposition, and the cause of gravity\cite{Unif1,Unif1Vacdiel,Unif1Grv,Unif1D-GEqn}.  Overall experimental observations suggest that \cite{Unif1,Unif1Vac}  the vacuum is filled of electrically neutral but polarisable building entities, called vacuuons (Fig.  \ref{fig2}), each composed of a spinning charge $+e$ at the core and $-e$ on the concentric spherical shell   bound strongly each other by a Coulomb force, and of spin angular momenta  $\frac{1}{2}\hbar$ and $-\frac{1}{2}\hbar$. 
This  vacuum will be polarised about an external charge,  building an electrostatic potential in which, in the cage formed by neighboring vacuuons,  the charge in turn maintains its oscillation.
Ordinarily  the vacuuons have each an external-effective spin $-\frac{1}{2}\hbar$  but are opposite aligned with their  closest neighbours
 and yet  in an applied magnetic field
some of the pairs will be broken into parallel aligned;  this vacuum is a magnetically susceptible paramagnetic.

Such a  vacuum is in particular pointed to  by  the observational pair processes taking place at the matter-vacuum interface. In
 the annihilation of a free electron $e^{\min}$ and positron $e^+$ at rest,
 $e^{\min}+e^+ \rightarrow 2 \g$, for example,    the  energy of the radiated  two $\g$'s, $2\times 511$ keV (see Refs.  in Sec. \Secexp.3), is  converted from {\it only}  the rest masses of $e^{\min},e^+$, whereas the Coulomb potential   between their charges at a separation distance $r_{+\min}$,
 $V_{+\min}=-e^2/4\pi\epsilon_0 r_{+\min}$ is not released, nor is the energy of the spins. Energy conservation requires that, after annihilation the  $V_{+\min}$ as well as the spin energy must  be conveyed  by a certain entity (the vacuuon here)   in the vacuum,  a point therein yet observationally  no different from any other points, whence the  vacuuonic vacuum.

\section*{3. Solutions for IED particle: an outline }
\def\secsolu{3}
          Sections  3.1-3.8
outline some of the  key solutions  obtained for the IED particle in \cite{Unif1,Unif1Schr,Unif1dBw,Unif1mass,Unif1Radiation,Unif1Vac,Unif1Vacdiel,Unif1Grv,Unif1D-GEqn,Unif1Dirac}.

{\it \secsolu.1 }
 Let for illustration  a given oscillatory charge $q$ of a characteristic frequency $\W$ be in contact with a linear chain of the vacuum along the $X$- axis, and be  oscillating along the $Z$-axis about an equilibrium site which moves at velocity $\vel$ in $+X$-direction.  The charge thereby generates two opposite travelling electromagnetic waves
in the $+X$-,$-X$-directions, given from solving (\ref{eq-maxwel1})  in dimensionless functions as:
 $\varphi^{\dagsup}=C_1e^{i[k^{\dagsup}x-\w^{\dagsup} t +\a_0]}, $
$\varphi^{\ddagsup}=-C_1e^{ i[k^{\ddagsup}x+\w^{\ddagsup} t -\a_0]} $
(Fig. \ref{fig1}b)
with $\vphi^j=|\Eb^j|/E_q$,  $E_q$  the amplitude of $E^j$, 
$x=X-X_q$,
$\w^j,k^j $ the Doppler displaced values of $\W,K$ as of (\ref{eq-Doppler}), and $\a_0$ the initial phase;
$\vphi^j$ is as with $\Eb^j$ a transverse wave displacement in coordinate space along the $Z$-axis.
The $\vphi^j$'s and the charge $q$ make up our IED particle; it has a total wave  \cite{Unif1,Unif1Schr,Unif1dBw}
$$ \displaylines{
\refstepcounter{equation} \label{eq-dBw1aa}
\psi =\varphi^{\dagsup}+\varphi^{\dagsup}=\phi  f, \quad \phi=e^{i[(K +\frac{\vel}{c}k_d ) x-\frac{\vel}{c}\w t]},
 \quad f = Ce^{i (k_d x-\w t+\a_0) }, \quad {\rm where } \hfill (\ref{eq-dBw1aa})
\cr
k_d
=(k^\dagsup-k^\ddagsup) /2=\sqrt{(k^\dagsup-K)(K-k^\ddagsup)}=\g K_d, \quad K_d =(\vel/c) K, \hfill (\ref{eq-dBw1aa}a)
\cr
\w= (\w^\dagsup+\w^\ddagsup)/2 =\sqrt{\w^\dagsup \w^\ddagsup}
=\g \W, \quad \g=\sqrt{\g^\dagsup\g^\ddagsup}=1/\sqrt{1-\vel^2/c^2}; \quad {\rm and} \hfill(\ref{eq-dBw1aa}b)
\cr
\lim_{K>>k_d }\phi =1, \quad  \lim_{K>>k_d} \psi=f;  \quad  \Psim=\lim_{\vel^2/c^2\rightarrow 0, K>>k_d}\psi= Ce^{i (K_d x-\Wvel t+\a_0) },
\hfill (\ref{eq-dBw1aa})'
}$$
with $C=4C_1=1/\sqrt{L}$ from normalisation of $\psi$, and $\Wvel $ as expressed after (\ref{eq-dB1a}) later. From its functional in (\ref{eq-dBw1aa}) and the wave and dynamical variable relations (\ref{eq-dB1a}) below, it follows   that \cite{Unif1Schr,Unif1dBw}   $\psi $  is equivalent to  the {\it de Broglie phase wave} \cite{deBroglie},
with $f$ the modulation envelope (dot-dashed line in Fig \ref{fig1}b);
  $k_d$  thus is    the de Broglie wavevector and $\lam_d=2\pi/k_d$ wavelength,  and $\w$  is the total  frequency. The $\vel^2/c^2\rightarrow 0$ limit of $\psi$,  $\Psim$, identifies with the Schr\"odinger wave function for a corresponding free particle.
From  (\ref{eq-dBw1aa}) further follows that
$\psi$ travels at a phase velocity  $W_p=\w/k_d=c^2/\vel$ and group velocity $W_g=(\w^{\dagsup}-\w^{\ddagsup})/[k^{\dagsup}-(-k^{\ddagsup})]
\dot{=}\vel $;  the particle's total energy $\eng$ and mass $m$ each travel at the velocity $W_g$ or $\vel$ (elaborated in updated edition of \cite{Unif1Schr}, internal).

{\it \secsolu.2 }
Following classical electrodynamics
 the electromagnetic waves have at every location $X$  a (mean)  energy density  $\eng_1(=\sqrt{\eng_1^{\dagsup}\eng_1^{\ddagsup}})=\epsilon_0 E_q^2 $ and linear momentum density $p_1=\eng_1/c$.
For our applications  here   in general $\eng_1$ is significantly lesser than  the total energy  $ \eng_q$ of the charge which  can thus without  "refuel" 
  oscillate continuously for a finite time, 
generating wave  trains of a (mean) total length $\Lw >>2\pi/K$.
The particle is as in  reality inevitably situated between some massive  walls say spaced at distance $L$ (Fig. \ref{fig1});
its wave amplitude is thereby quantized as $E_q^2=n_{ph}E_{q.ph}^2 $, with   $ n_{ph}=1, 2, \ldots$, given as a direct solution for the charge in harmonic oscillation.
The total wave energy and linear momentum of the wave train thus are
$
\eng (=\eng_q)=\Lw \eng_1=   n_{ph}  \eng_{ph} $, with $\eng_{ph}= \Lw \epsilon_0 E_{q.ph}^2 $ an energy quntum; 
the explicit value of $\eng_{ph} $ follows from (\ref{eq-Plancke}), law (b), to be  $\eng_{ph}=\hbar \w$.
We shall below refer to  the single charged electron, proton etc. for which  $n_{ph}=1$ according to experiments; so $\eng=\hbar \w$.
In all, the above depicts the IED process in the established unified framework of  classical and quantum electrodynamics.

  $\Eng $ $=\lim_{\vel^2/c^2\rightarrow 0}\eng =\hbar \W$ gives    the ground state of a smallest quantum; thus $\Eng$ can  not be dissipated  except in a pair annihilation.
The charge repeatedly re-absorbs the radiation on reflection from the walls and then re-emits, maintaining therefore $\eng$ constant.
The re-absorption of reflected waves $\vphi^j$'s, thus  $\psi$, by the charge $q$  is further  ensured by: (i) The $\vphi^j$'s are in natural resonance with the source.
(ii) At (non-annihilating) massive walls,  irrespective  of the incident angle the  $\vphi^j$'s, thus $\psi$,  as a whole will always be {\it reciprocally} reflected to the $q$,  via an usual  "temporary absorption and emission" scheme but  here by a {\it vacuuon}; the waves are of too high frequencies $\w^j$'s   to be   absorbed  by a material particle in the wall. The vacuuon invariably  is polarised in the static field of $q$,  thus bound to the charges in the massive wall and scatters the waves reciprocally on conserving total momentum.

Subtracting the total rest energy and quadratic rest linear momentum  from the relativistic ones gives  the kinetic energy and linear momentum of the particle
 $\eng_\vel(\equiv \frac{1}{2}m\vel^2)
=\hbar (\w-\W)$, $p_\vel (\equiv m \vel)=\sqrt{(\hbar k)^2-(\hbar K)^2}$.
With $\g$ $=1+\frac{1}{2}\frac{\vel^2}{c^2}+\frac{3}{8}\frac{\vel^4}{c^4}+\ldots$, $\g^2-1=(\vel/c)^2\g^2$, reorganising, these reduce to the usual form of {\it de Broglie relations}
$$ \displaylines{\refstepcounter{equation} \label{eq-engvelm1}\label{eq-dB1a}
\eng_\vel
=\hbar \wvel,
\quad p_\vel= 2\pi\hbar /\lam_d
\hfill (\ref{eq-dB1a})
}$$
where
$\wvel=\g' \Wvel$, $\g'=1+(3/4)(\vel^2/c^2)+\ldots$,
$ \Wvel=(1/2)(\vel^2/c^2)\W$,
$\lam_d=2\pi/k_d$.

{\it \secsolu.3 }
More generally,  the Maxwell's equations (\ref{eq-maxwel1}) in an applied potential $V_{a}$ field lead  to  a  wave equation for the total wave $\psi$ $(c^2 +V_a/m)\nabla^2\psi =\pd^2_t\psi$; this at the limits $\vel^2/c^2\rightarrow 0$ and $K>> K_d$ reduces  to an equation governing directly the particle's kinetic motion\cite{Unif1Schr}  which is  equivalent to the {\it Schr\"odinger equation}\cite{Schrodinger},
$$ \displaylines{\refstepcounter{equation}\label{eq-sch}
H\Psim =i\hbar \pd_t \Psim,
\quad {\rm where }\ H= -(1/2M)\nabla^2 +V_a.  \hfill (\ref{eq-sch})
}$$
For two spin half  IED particles in region $\vel^2/c^2>0$, (\ref{eq-maxwel1}) lead to a {\it Dirac equation}\cite{Unif1Dirac}.

{\it \secsolu.4 }
If an IED particle
moving at velocity $\vel$ is decelerated to say at rest in the vacuum, then its total wave  $\psi$ of (\ref{eq-dBw1aa}) deconvolutes off a thermal mode  $k_d$ into $\psi_0=e^{i[K x -\W t]}$,   $\psi$ and $\psi_0$ being each the totals of electromagnetic waves regularly comprising the particle in the respective normal states. The deconvolution  is an inverse process of the de Broglie wave formation (Sec. 3.1). The oscillation at mode $k_d$ of the deconvoluted de Broglie wave  acts as an apparent source,  generating   an electromagnetic wave of wavevector $k_{rad}= K_d (c/\vel)$; this gives a {\it thermal radiation}\cite{Unif1Radiation};
and conversely, a {\it thermal absorption}.

{\it \secsolu.5 }
Whereas equations (\ref{eq-dBw1aa})-(\ref{eq-sch}) convey all the essential {\it wave attributes}, these together  with the point charge $q$  convey also all the essential {\it point-like attributes} of a particle as observationally known in three basic ways:
(i) The  $\eng_\vel$,$p_\vel$ contain all the information on the linear  dynamics known with  a  point object.
(ii) An IED particle would  interact with a detector (e.g. by absorption), or another particle, through its extensive $\psi$ at a {\it fixed interface} or through its {\it point charge},   each  manifesting a spatially  point event.
(iii) In a condensed matter, each particle (a nuclei, electron, atom, etc.) will be anchored through  its  point  charge, as a mass center, about a fixed position or in a finite region; its  waves would typically be confined to a region   by reflection from neighbouring particles or by moving in a closed path.

{\it \secsolu.6 }
The wave trains of $\vphi^j$'s   in rectilinear motion at the speed of light $c(=\w^j/k^j)$  resemble each a  {\it rigid} object and thus obey Newtonian mechanics; given $c$ is {\it finite}  instead of  infinite, the wave trains must have a  {\it finite} (mean) inertial mass ($m$) instead of zero. From these  and the relation $\eng=pc$ earlier, with $p,\eng$ being now the {\it linear momentum} and {\it kinetic energy} of the {\it rigid wave train},  follows  (\ref{eq-plnkmass1a}) below (Newton's law of inertia); and (\ref{eq-plnkmass1a}), (\ref{eq-Plancke}) and (\ref{eq-dBw1aa}b) further give   (\ref{eq-rel-mc2bb}) below \cite{Unif1,Unif1Schr,Unif1mass}:
 $$\displaylines{
\refstepcounter{equation} \label{eq-plnkmass1a}
p=mc, \quad \eng=m c^2; \quad M^2 c^4+p_\vel^2c^2 =\eng^2;
\hfill (\ref{eq-plnkmass1a})
\cr
\refstepcounter{equation} \label{eq-rel-mc2bb}
m=\g M, \quad   M=\hbar \W/c^2 \ \ \mbox{(for $n_{ph}=1$) }
\hfill (\ref{eq-rel-mc2bb})
}$$
with $m$ the relativistic and $M$ the {\it rest mass} of the wave train and thus of the particle, noting that rest mass is intrinsic of an object irrespective of in which  motion the object is in.

\begin{figure}[h] \includegraphics[width=0.55\textwidth]{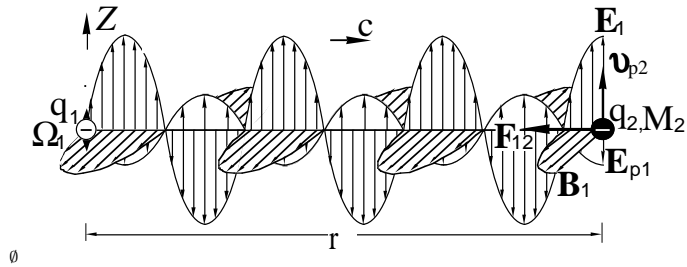}
\vspace{-0.5cm}
 \caption{Particle 2 of charge $q_2$ and mass $M_2$ is in the fields $\Eb_{p1},\Bb_1$ of  charge $q_1$ of particle 1, oscillating with  frequency $\W_1$, acted by an attractive Lorentz force $\Fb_{12}$.  }\label{fig-grv} \end{figure}

{\it \secsolu.7 }
Two IED particles of masses $M_1,M_2$ ($=\frac{\hbar \W_1}{c^2}, \frac{\hbar \W_2}{c^2}$) and charges $q_1,q_2$, $=\pm e$ here, separated at $r$ apart  in a paramagnetic dielectric vacuum are always  attracted one another by a Lorentz or attractive radiation force (Fig. \ref{fig-grv}). 
This force acting on charge $\ip$ at time $T$, in charge $i$'s depolarisation
 field $\Eb_{pi}=-\chi_{_{\v e}}\Eb_i$ and
 magnetic field $\Bb_i $ is
$\Fb_{i\ip}(r,T)=q_\ip \velb_{p\ip}(r,T) \times \Bb_{i}(r,T)$;
 $F_{i\ip\vphi}= \int^{\Lw{_i} /c}_{0} F_{i\ip} d T$ gives the total force due particle $i$ of  a wave train length $L_{\vphi i}$.
Here, $\velb_{p\ip}=\int q_\ip \Eb_{pi}/M_\ip d T$;
 $\chi_{_{\v e}}$,$\chi_{_{\v m}}$ are the electric and magnetic susceptibilities of the vacuum; $\Eb_i=\sqrt{\Eb_i^{\dagsup}\Eb_i^{\ddagsup} }$, etc.;
$\Bb_i= \Bb_{\empty i} + \Bb_{mi}$, with $\Bb_{\empty i}$ applied  in empty space and $ \Bb_{mi}=\frac{\chi_{_{\v m}}}{\chi_{_{\v m}}+1}\Bb_{i}$ induced in vacuum; $i,\ip=1,2$.
 The matter-penetrating (due to $\Bb_{mi},\Bb_{m\ip}$) mutual mean   attractive radiation force   is (JXZJ, internal; \cite{Unif1Grv}):
 $$\displaylines{\refstepcounter{equation} \label{eq-grv}
F=\sqrt{F_{12 \vphi} F_{21 \vphi}}= G M_1M_2/r^2, \quad  {\rm where} \ G=\zeta \chi_{_{\v m}} \chi_{_{\v e}} e^4/(\chi_{_{\v m}}+1) \epsilon_0^2 h^2 \rho_l.
 \hfill (\ref{eq-grv})
}$$
  $\rho_l$ is  the linear mass density of vacuum;     $\zeta $ is a numerical constant depending on the averaging method, $\zeta =\pi $ given  in \cite{Unif1Grv} (2006) and is being refined.
$F$ is an attraction irrespective of the sings of the charges, is not shielded by matter as the underlying vacuuon dipole- and spin- waves are not,   and has an inverse square formula; this $F$ resembles in all respects {\it Newton's gravitational force}.

\begin{figure}[htpb] \includegraphics[width=0.55\textwidth]{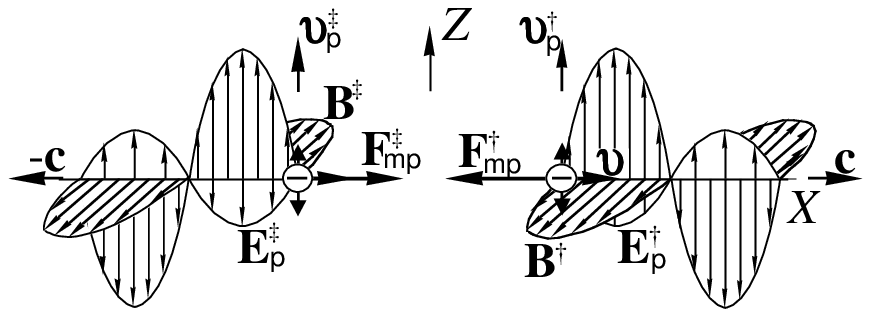}
\vspace{-0.8cm}
 \caption{A  particle  of charge $q$ and mass $ M$ is in the fields $\Eb_{p}^j,\Bb^j$ induced by its own $q$   acted by an attractive Lorentz force $\Fb_{mp}^{\dagsup}-\Fb_{mp}^{\ddagsup}=\Fb_{mp}$ opposing its motion of velocity $\vel$. }\label{fig-DG} \end{figure}

{\it \secsolu.8 }
Similar to the gravity in  Sec.  \secsolu.7,  an IED  particle is always attracted by a Lorentz force $F^j$ (Fig. \ref{fig-DG})  acting on its own charge $q$ in the $E^j_{p}$,$B^j$  fields induced by $q$ itself in a dielectric medium; $j=\dagger,\ddagger$. The net force  $F^{\dagsup}-F^{\ddagsup}$ presents a frictional  force \cite{Unif1D-GEqn}   $f=(b_1/L_\vphi \psitot)d \psitot/d T$ opposing the particle's  motion, with $b_1$  a constant of the medium and charge $q$; when the medium identifies with a dielectric vacuum, $f$ depicts a {\it self gravity} on the particle. 
The wave equation in the presence of  $f$ \cite{Unif1D-GEqn}  is equivalent to the {\it Doebner-Goldin equation} predicted by H.-D. Doebner and G.A. Goldin in group theoretical terms\cite{D-G}:
$$ \displaylines{\refstepcounter{equation}\label{eq-schDG}
(H+iD\hbar G)\Psim =i\hbar \pd_t \Psim,
\quad \ G=\nabla^2 \Psim +|\nabla \Psim|^2/|\Psim|^2, \hfill (\ref{eq-schDG})
}$$
with $D$ depending on $f$, and $H$ the usual Hamiltonian operator as in (\ref{eq-sch}). (\ref{eq-schDG}) precedes a "grand unified"  {\it wave equation including gravity} between particles.

\vspace{-0.4cm}
\section*{4. Validating the IED model: the solutions and experiments} 
Equations (\ref{eq-dBw1aa})--(\ref{eq-schDG}) are exact predictions of a set of familiar basic equations  of particles in contemporary  physics, originally proposed by several individual physicists  on hypothetical  or phenomenological basis;  (\ref{eq-dBw1aa})--(\ref{eq-grv}) and  the associated properties (Secs.\secsolu,\Secexp) have long been broadly experimentally  corroborated or demonstrated. We below review the key experiments, and   underline their specific indications of the IED model and also the insufficiency of  otherwise pictures if in question.

%
%

\begin{figure}[htpb] \includegraphics[width=0.55\textwidth]{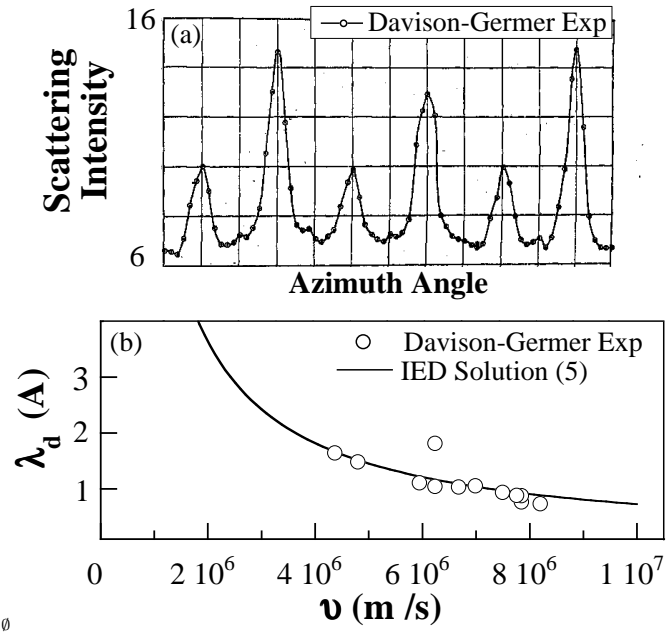}
\vspace{-0.5cm}
\caption{  (a)  Diffraction fringe intensity v.s. azimuth angle ($90^o-\theta$), and  (b) de Broglie wavelength $\lam_d$ v.s. velocity $\vel$, circles, for electrons diffracted from a  crystal grating measured in \cite{Davisson-Germer1927}. Solid line in (a) is after the  IED  solutions for de Broglie relations given in    (\ref{eq-dB1a}).}\label{fig-Davison-Germer-dif}\label{fig-Davison-Germer}
\end{figure}

\begin{figure}[ht] \includegraphics[width=0.55\textwidth]{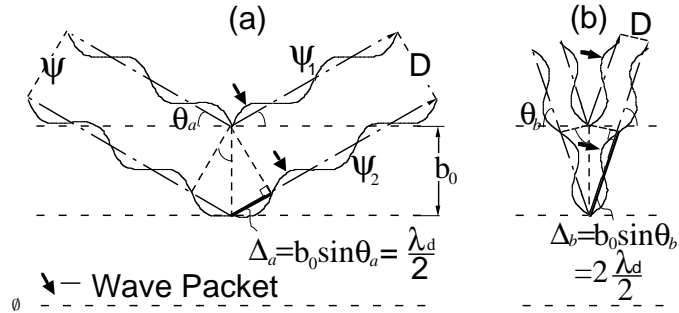}
 \caption{An electron plane wave   $\psi (\approx f)$ of wavelength $\lam_d$ diffracted from two adjacent planes differ by (a)  $\D_a$ for incident angle $\theta_a$, and (b) $\D_b$ for $\theta_b$,  each producing  diffraction peaks. Two  wave packets (indicated by thick arrows), if  superposing to a peak at angle $\theta_a$ in (a), will not do so at angle $\theta_b$ in (b).} \label{fig-Davison-Germer-dev} \end{figure}

{\it \Secexp.1 }
A wave characteristic of the material particles as of (\ref{eq-dBw1aa})-(\ref{eq-sch}) is  broadly experimentally established today for  electrons\cite{Davisson-Germer1927,GPThomsom1927}, atoms and molecules \cite{EstermannStern1930}, neutrons\cite{HalbanPreeiswerk1936,MitchellPowers1936,Elsasser1936}, and large molecules\cite{Brezger2002}.
 In the first historical experimental demonstrations \cite{Davisson-Germer1927,GPThomsom1927},  electrons  of well controlled  kinetic energy ($\eng_\vel$) were let  stricken 
 on to a crystal at angle $\theta$ from its planes spaced at $b_0$. These produced diffraction fringes (Fig. \ref{fig-Davison-Germer-dif}a) according to (i)  Bragg formula $2b_0\sin \theta=n\lam_d$ (cf. Fig. \ref{fig-Davison-Germer-dev}) in the same way as the light waves and  ordinary  elastic  waves do,  and (ii) the de Broglie relations as of  (\ref{eq-dB1a}),  $\lam_d=h/\sqrt{2m_e \eng_\vel}$ (solid line, Fig. \ref{fig-Davison-Germer-dif}b). For the high velocity ($\vel\sim 9.4 \times 10^7$ m/s) electrons used in \cite{GPThomsom1927}, a relativistic $\lam_d(=\g 2\pi/K_d)$ was obtained, indicating a full wave function as of $\psi$ in (\ref{eq-dBw1aa}),  $\Psim$  being thus its  $\vel^2/c^2\rightarrow 0$ limit.

 Although the stationary  wave $\psi_i$ will  upon detection  (at D in Fig \ref{fig-Davison-Germer-dev}) be generally "collapsed" by e.g. emitting radiation, however the coherent interference producing the diffraction fringes can only have occurred before the collapsing and between two  stationary-state  electron waves  $\psi_i$'s, Fig. \ref{fig-Davison-Germer-dev}.
Therefore the diffraction fringes inform  that a wave  $\psi$ as of (\ref{eq-dBw1aa}) presents regularly with a stationary-state electron.

The diffraction fringes  need necessarily be produced by the interference between two {\it travelling plane waves}, as illustrated in Fig. \ref{fig-Davison-Germer-dev} a,b for two Bragg diffractions at angles $\theta_a,\theta _b$. The IED particle at  scale $k_d$, the $f$ or $\Psim$ of  (\ref{eq-dBw1aa})$'$,  is a travelling plane wave  in a self-sufficient way; its "particle" attribute is facilitated by the IED model  itself (Sec. 3.5). Alternatively, as a practical means  today, a "particle" attribute is attached  to the  Schr\"odinger plane wave by dispersing  (supposing a physical basis exists) it into a wave packet; the latter is no longer a plane wave and,  as shown by the thick arrows in Fig. \ref{fig-Davison-Germer-dev}a,b, would not produce  diffraction fringes.

{\it \Secexp.2 }  Experiments for electrons using certain kinds of a double slit 
 since the1970's \cite{Merli-etal-1976,Tonomura:1989}, and for neutrons from the very first experiments using crystal diffraction  \cite{HalbanPreeiswerk1936,MitchellPowers1936}  and using double slit, as a hitherto  most precise realisation for matter waves,    in 1988\cite{Zeilingge1988} 
as judged by  the generally low neutron flux intensity  \cite{Zeilingge1988,Rauch-Werner2000}, have shown that, the interference pattern
\begin{figure}[bhtp]
\includegraphics[width=0.3\textwidth]{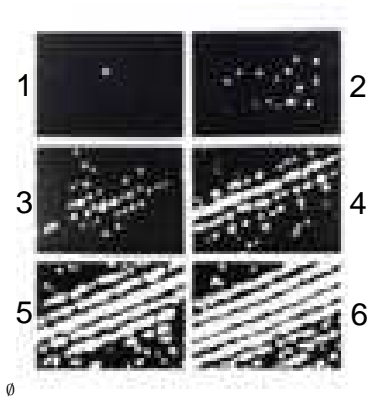}
\vspace{-0.5cm}
\caption{Experimental  single electron self interference fringe in a double-slit type of  device TV filmed in \cite{Merli-etal-1976};  1,2 \ldots,6 indicate  increasing exposure time.   }\label{fig-Bolony-diff} \end{figure}
 (as in Fig. \ref{fig-Bolony-diff} \cite{Merli-etal-1976}) is just as well  produced when only one particle passes through an interferometer at a time. That is, {\it each particle  interferes with itself}.

\begin{figure}[htpb]
\includegraphics[width=0.55\textwidth]{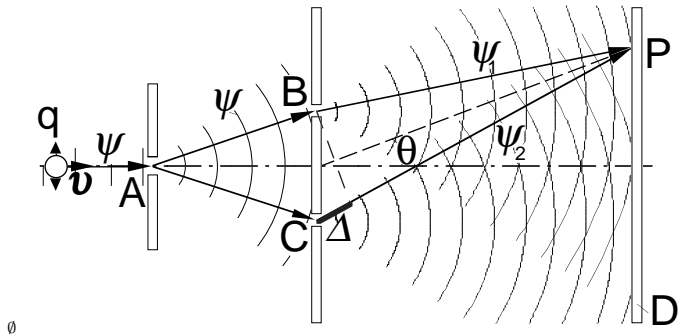}
\vspace{-0.4cm}
\caption{ Schematic illustration of a single IED electron $\psi$ self interference  in a double slit.  }  \label{fig-double-slitGrp27} \end{figure}

Self interference, such as in a double slit (Fig.  \ref{fig-double-slitGrp27})  requires each particle passes two slits at the {\it same} time. ---For a statistical point particle this is a logical impossibility.  
The IED particle ($\psi$) naturally has this ability  since its each constituent electromagnetic wave ($\vphi^j$), hence the total $\psi$, will in an open vacuum medium  disseminate itself in all possible directions, which is based on observational fact and also the understood principle for  medium waves.
As illustrated in Fig. \ref{fig-double-slitGrp27},  the  total wave $\psi$ of an IED particle arriving at  narrow slit $A$  will be regenerated, by Huygens' principle or as solution to (\ref{eq-maxwel1}) in three dimensions, as a spherical  wave  $\psi(\rb,T)$; two of the partial components $\psi(AB,T), \psi(AC,T)$ along equidistance paths 
$AB,AC$ 
will enter the two slits $B,C$ at the same time. The  resultant two split waves $\psi_1(\rb_1,T)$ and $\psi_2(\rb_2,T)$ will traverse distances $\rb_1=BP$ and $ \rb_2=CP$,  rejoin at  point $P$ on a photographic plate $D$ as $\psi_{tot}=\psi_1+\psi_2$,  and will superpose with constructive interference into a peak if $\Delta=r_1-r_2=n \lam_d$, with
 $n$ an  integer. Or else they annul each other. If a thermal mode of the  $\psi_{tot}$ peak is absorbed by a molecule at $P$ in $D$, a detection (excitation) signal will  be produced without the arriving of $q$.  The charge $q$,  if finally arriving at detector $D$ too, is propelled  forward at  velocity $\vel$ by  a repeated re-absorption/re-emission of  $\psi$ travelling at the enormous phase velocity $c^2/\vel, >>\vel$, by Sec. 3.1. The so driven charge will definitely first travel to  $A$,  then statistically take a radial path, say $AB$, and on exiting $B$, continue along BP only if P is a diffraction peak which feeds the charge with  a linear momentum $k_d$. Or else, the charge  gets no feed of $k_d$ and will stray  off the course (detailed treatment given in internal report). The self interference is one of the critical tests that point to  the IED model is  not just sufficient but also  necessary.

In theory, interference is understood to be  the (only feasible) result of superposition of  vector fields, here the $\Eb_i$'s or $\psi_i$'s $(=E_i/E_q)$, at any point (or oscillator)  in the medium; in contrast, two identical fermionic particles as a whole tend to repel each other (Pauli principle). This tends to suggest that, even in a many-particle beam as in \cite{Davisson-Germer1927,GPThomsom1927}, diffraction is predominantly the result of self-interference of each individual particle in an afore-discussed fashion.

{\it \Secexp.3 }
The various pair production and annihilation experiments of   elementary particles provide a most direct revelation that, apart from charges, electromagnetic waves actually constitute  the material particles  as stated by  the IED model.
In the same example as in Sec. \SecIEDmodel, an electron $e^{\minus}$ and   positron $e^{\p}$ \cite{Dirac1928,Anderson1932,Blackett1933}
can annihilate  into (typically  in a condensed matter) two  gamma rays $\g$'s,  $ \el^{\minus} +\el^{+}\rightarrow 2\gamma$,
with the two $\g$'s being  emitted \cite{Thribaud1934,Joliot1933,Klemperer1934,Yoshizawa1984}  in opposite directions and
 carrying a total  energy
$\eng_{2\g}=2\hbar \w_{r0} =2\times (511.0031\pm 0.0032)  $ keV  (precision Ge(Li)- scintillation data from \cite{Yoshizawa1984}).  This $\eng_{2\g}$ value equals twice the electron rest mass  $M_e$ times $ c^2$, $2\times M_e c^2=2\times 510.9989$ keV (CODATA, 1998), with which the IED solution (\ref{eq-plnkmass1a}) directly agrees.

\begin{figure}[h] \includegraphics[width=0.3\textwidth]{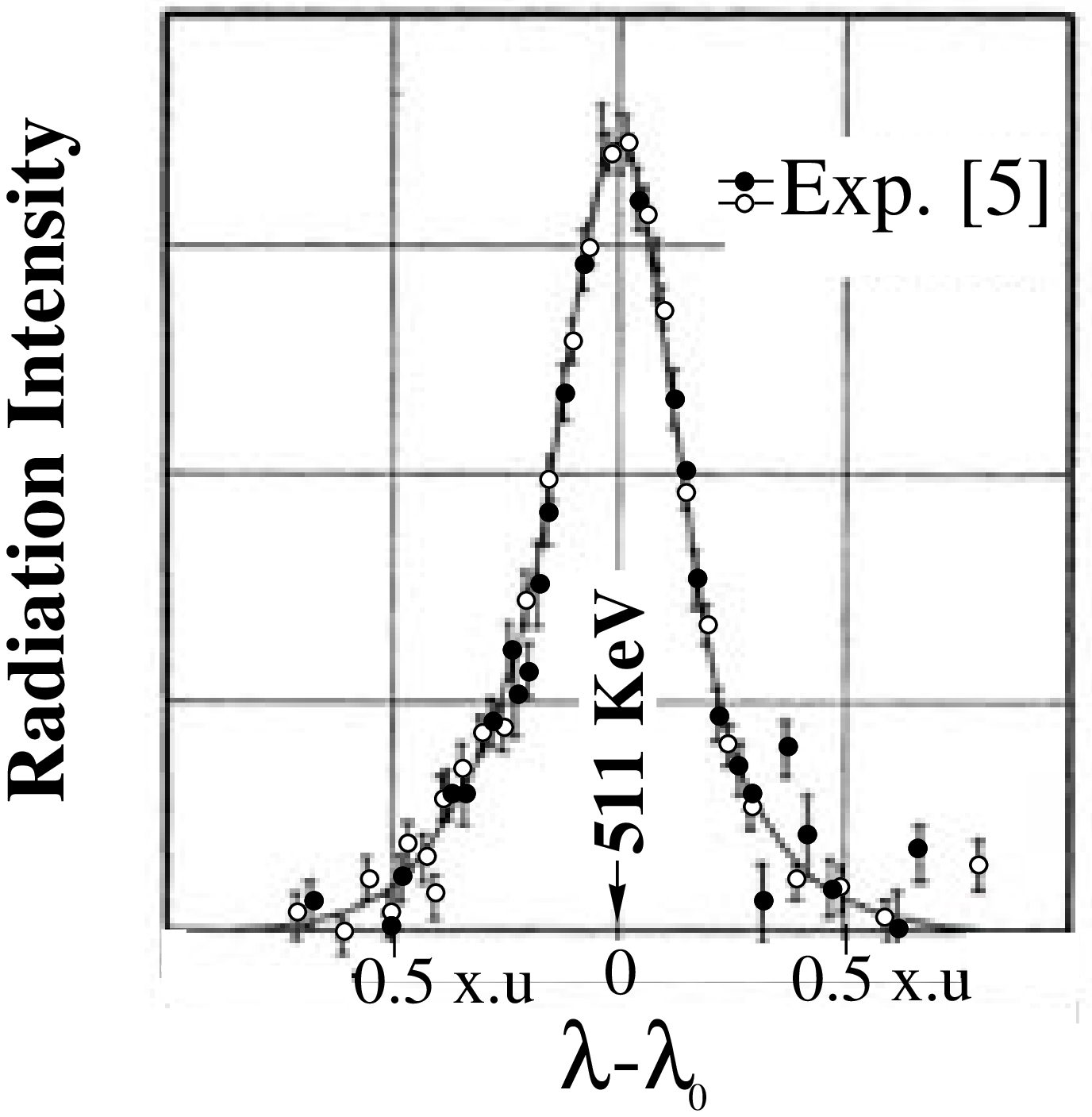}
\vspace{-0.5cm}
\caption{Experimental $e^{\min},e^+$ annihilation radiation intensity v.s.  wavelength displacement\cite{DuMond1949}.} \label{fig7-DuMond1949}\end{figure}

When emitting the radiation, the annihilating  particles $e^{\min},e^+$ are not  undergoing any accelerations inasmuch as is externally observable; rather, most favourably they are at {\it rest} as indicated by the  peak position  at 511 keV in
Fig. \ref{fig7-DuMond1949}. 
However, electromagnetic theory requires the charge must undergo  {\it acceleration} ($>$ or $<0$) to emit  radiation. Inevitably then, the emitted $\g$- electromagnetic waves ought to have been as stated by the IED model regularly generated  by the oscillation of charge {\it within} each normal-state particle; and upon annihilation these waves are no longer re-absorbed by the charges which have now neutralised one another into the vacuum.

{\it \Secexp.4 }
If the particles  $e^{\min},e^+$ are during annihilation  in motion, then according to the IED model  their constituting electromagnetic waves should be  subject to  a Doppler effect, law (c). This effect indeed is
directly revealed by
the commonly observed unusual broadening in the annihilation radiation intensity profile, as shown in Fig. \ref{fig7-DuMond1949} as a function of differential wavelength $\lam-\lam_0$ measured in \cite{DuMond1949} using a crystal spectrometer, $\lam_0=h/M_e c$ being due to   electron rest mass $M_e$. It was   historically first shown in  \cite{DuMond1949} that a  residual broadening $\delta (\lam-\lam_0)$ $\sim$0.096 x.u.  ($\sim 0.0958$ A) retained after subtracting  an instrumental cause and was identified   to result from the Doppler effect from the thermal motion, of a velocity $\sim$16 eV, of the recombining positron-(conduction) electron pair in the metal specimen.

\begin{figure}[htpb] \includegraphics[width=0.336\textwidth]{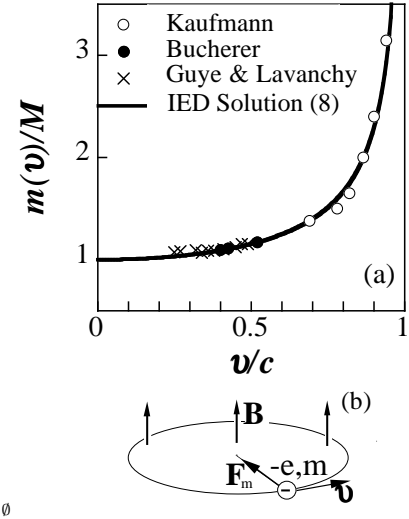}
\vspace{-0.5cm}
\caption{Experimental  mass v.s. velocity of electron (a) in a  set-up as  (b)\cite{Kaufmann1906,Bucherer1908,Guye:1915}.}   \label{fig-mass-v-Kaufmanexp}\end{figure}

{\it \Secexp.5 } The IED solution (\ref{eq-rel-mc2bb}) for mass   v.s. velocity, $m=\g M$ (solid curve, Fig. \ref{fig-mass-v-Kaufmanexp}a) predicts exactly the well-known empirical formula concluded originally from a series of experiments on electrons during 1901-15 [26-28].
There,  electrons are driven into motion at a given velocity ($\vel$) in an applied magnetic field ($\Bb$) perpendicular to the electron motion plane  (Fig \ref{fig-mass-v-Kaufmanexp}b), and are thus   subject to a Lorentz force ${\bf F}_{\rm{m}}=-e {\pmb \vel}\times \Bb $; the equation of motion is $F_{\rm{m}}=m \vel^2/R(y,z)$. A measurement of their coordinates $y,z$    gives then  an {\it in situ}, non-destructive determination of  the $e/m$ ratio, hence the $m$ value, v.s. $\vel$ (circles, Fig \ref{fig-mass-v-Kaufmanexp}a).

 The exact prediction by  (\ref{eq-rel-mc2bb})
of the experimental relation $m=\g M$  points to an unique underling IED  process in two major ways: (i) From its relation to  total energy ($\eng$),
 the IED represents a unique  feasible   {\it microscopic process} (Sec. \secsolu.6) for yielding the exact solution relation (\ref{eq-plnkmass1a}),  $\eng=mc^2$, where the $\vel$-dependence of $\eng$, hence $m$, results automatically  from this process owing to a Doppler effect which separately is an experimental law, (c). The relation $\eng=mc^2$ originally was  a postulate by A Einstein (1905), being aware of the experiments of \cite{Kaufmann1906}, and up to  the present  it has not been understood why total energy $\eng$ equals mass $m$ times the square of the velocity of light $c^2$. (ii) Observationally\cite{Kaufmann1906,Bucherer1908,Guye:1915,Elderetal1947}, an electron in macroscopic uniform circular motion emits radiation statistically, so  sometimes  it appears not radiating; also, an orbiting atomic-electron in a stationary state does not radiate. The above thus appears paradoxical to the electromagnetic theory by which  the charge under constant centripetal acceleration  ($a_R$) should be constantly radiating. This paradox  presents not with an  IED electron whose  oscillatory charge emits electromagnetic radiation ($\vphi^j$'s) {\it all the time} and  the acceleration $a_R$ additionally bends its
radiation  wave path  from linear into  circular. It is self-evident that, the radiation as emitted by a circulating electron is {\it not} due to the centripetal acceleration, but rather is due to the sudden deceleration associated with a mass reduction,  $(m-M)c^2$, accompanying which the electron falls onto a smaller-radius orbit if no energy compensation.

{\it \Secexp.6 }
The inverse square formula of gravity as of (\ref{eq-grv}) was originally an empirical law discovered  by I Newton (1687) based on the then available observational astronomical data. Since the first "torsion balance"  determination of $G$ value  by H Cavendish  (1798), the constancy of  $G$ and  the inverse square behaviour of $F$ have been repeatedly  experimentally (terrestrially) validated with ever improving  accuracy \cite{Gillies1997}. The cause of gravity is up to the present yet not understood. Gravity, as a fundamental force  universally exerted between all masses at all distances, is as intrinsic a property as the wave function, mass and charge of a particle, so that a particle model could not be said to be adequate  without a built-in scheme for this force, the gravity. The IED model uniquely has such a scheme and predicts a precise inverse square formula (\ref{eq-grv}),  with its constant $G$ being expressed by only fundamental constants and constants of the vacuum.

\section*{5. Concluding remarks}

A set of well known basic equations embedding our present-day essential  knowledge of particles can be derived based on  solutions for the   IED process governed by a minimal set of long established basic or first principles laws (a-e). These equations and the closely related  particle properties have long  been broadly experimentally validated; the present  critical review of several key experiments suggests that the underling IED process is not just sufficient  but is also necessary. An otherwise particle picture would not yield  all of the predictions  or just some of these without some kind of clashes.

The author expresses thanks to Professor H.-D. Doebner for  valuable long discussions during/after recent conferences and  the author's visit to him in Munich (2008), which has motivated some of the  discussions (see further \cite{Unif1D-GEqn}) and also an immediate writing up of this review, to Professors H.-D. Doebner and G. Poghosyan  for giving the opportunity to present the research at the 27th Int Colloq Group Theo Meth in Phys,  Yerevan, 2008 (Group 27),  to Scientist P.-I. Johansson for  moral and funding support of the research,  to Scientists P.-I. Johansson and H. Rundlof  for first-hand  explanations about measurement issues relating to annihilation radiation and  neutron diffraction,  to Professor  G.A. Goldin for useful stimulating comments on the research and to
Professors M. d. Olmo, S.T. Ali,  G. Poghosyan, V.K. Dobrev, P. Bieliavsky, W. Zakrzewski,
T. Tchrakian,  W. Ruehl, I. Khabibullin, R. Dahm
for intellectual communications   at the Group 27, and to Dr. A. Gusev for helpful editorial
comments.


\begin{thebibliography}{10}
\bibitem{Unif1} J.  X.  Zheng-Johansson and P-I.  Johansson,
{\it Unification of Classical, Quantum and Relativistic Mechanics and of the Four Forces}    (Nova Sci.  Pub.  Inc., N. Y., 2006);
{\it Inference of Basic Laws of Classical, Quantum and Relativistic Mechanics from First-Principles Classical-Mechanics Solutions}   (Nova Sci.  Pub., Inc., N. Y., 2006).


\bibitem{Unif1Schr} J.  X.  Zheng-Johansson and P-I.  Johansson,
{Suppl.  Blug.  J.  Phys. }{\bf 33}, 763 (2006);
in {\it Quantum Theory and Symmetries} {\bf IV.2},
ed. by V.K. Dobrev (Heron Press, Sofia, 2006);
arxiv:phyiscs/0411134v5.

\bibitem{Unif1dBw} J.  X.  Zheng-Johansson and P-I.  Johansson, {Prog.  Phys. } {\bf 4},  32 (2006);
arxiv:phyiscs/0608265.

\bibitem{Unif1mass} J.  X.  Zheng-Johansson and P-I.  Johansson,
{Phys. Essays} {\bf 19}, 544 (2006);
arxiv:phyiscs/0501037.


\bibitem{Unif1Radiation}
J.  X.  Zheng-Johansson,
{Prog.   Phys. } {\bf 3}, 78 (2006);
arxiv:phyiscs/060616.


\bibitem{Unif1Dirac}
J.  X.  Zheng-Johansson,
{ J.  Phys.  Conf. Ser. {\bf 128}}, 012019 (2008);
     {\it Proc.  5th Int.  Symp. Quant.  Theory and Symmetries}, ed. by M. d. Olmo (Valladolid, 2007).

\bibitem{Unif1D-GEqn}
J.  X.  Zheng-Johansson,
arxiv:0801.4279.

\bibitem{Unif1Vacdiel}
J.  X.  Zheng-Johansson,
arxiv:physics/0612096.

\bibitem{Unif1Grv}
J.  X.  Zheng-Johansson  and P-I. Johansson,  
{Suppl.  Blug.  J.  Phys.} {\bf 33}, 771 (2006), with R. Lundin; \quad 
J.X.  Zheng-Johansson  and P-I. Johansson,  arxiv:phyiscs/0411245.


\bibitem{Unif1Vac}
J.  X.  Zheng-Johansson,
arxiv:0704.0131.

\bibitem{ParticleDataGroup2008} D.  Griffith, {\it Introduction to elementary particles} (Harper and Row Publisher, 1987);   L. Motanet {\it et al.}, {Phys.  Rev. } D {\bf 50}, 1173 (1994).


\bibitem{Planck:1900} M.  Planck, Ann.  Phys.  {\bf 1}, 69 (1900).

 \bibitem{Stark1905}J.  Stark,
{Z.  Phys. } {\bf 6}, 892 (1905);
{ Astr.  Phys.  J. } {\bf 25}, 23 (1907).

 \bibitem{DuMond1949}
J. W. M.  DuMond, D. A.  Lind, and B. B.  Watson, {  Phys.  Rev. } {\bf 75}, 1226 (1949).

%





\bibitem{deBroglie}
 L.  De Broglie,
{  Compt.  Rend. },   {\bf 177}, 507  (1923);
PhD Thesis, Univ.  of Paris, 1924;
     {  Phil.  Mag. } {\bf 47}, 446 (1924).




\bibitem{Schrodinger}
E.  Schr\"odinger,
{  Ann.  der Phys. }
{\bf 79}, 361 (1926);
{\it ibid. }  {\bf 79}, 489 (1926);
{\it ibid. }  {\bf 80}, 473 (1926);
{\it ibid. }  {\bf 81}, 109 (1926).

\bibitem{D-G} H. -D.  Doebner and G. A.  Goldin, {  Phys.  Lett. } {\bf A 162},  397 (1992); {  J.  Phys.  } A
{\bf 27}, 1771 (1994).

\bibitem{Davisson-Germer1927}
         C.  Davisson and L. H.  Germer,
{  Phys.  Rev. } {\bf 30}, 705 (1927);
{  Nature} {\bf 119}, 558(1927).
%


\bibitem{GPThomsom1927} G. P.  Thomson,
{  Roy.   Soc.  Proc. }  {\bf A117}, 600-609 (1927); 
G. P.  Thomson and A. Reid, {  Nature} {\bf 119}, 890 (1927).
%
%

\bibitem{EstermannStern1930} I.  Estermann and O.  Stern, Z.  Phys.  {\bf 61}, 95 (1930).


\bibitem{HalbanPreeiswerk1936} H.  Halban and P.  Preiswerk, {  Compt.  Rend. }
{\bf 203}, 73 (1936)

\bibitem{MitchellPowers1936} D. P.  Mitchell and P. N.  Powers, {  Phys.  Rev. } {\bf 50}, 486 (1936).


\bibitem{Elsasser1936} W.  M.  Elsasser,  {  Compt.  Rend.  }
{\bf 202}, 1029 (1936).

\bibitem{Brezger2002} B.  Brezger,
L.  Hackermller, S.  Uttenthaler, J.  Petschinka, M.  Arndt, A.  Zeilinger, {  Phys.  Rev.  Lett. } {\bf 88}, 100404 (2002).


\bibitem{Merli-etal-1976}
P. G.  Merli, G. F.  Missiroli, and G.  Pozzi,
{  Am.   J.  Phys. }  {\bf 44}, 306  (1976).




\bibitem{Tonomura:1989} A.  Tonomura {\it et al,}
{  Am.  J.  Phys. } {\bf 57}, 117 (1989).
                %
               %
               %




\bibitem{Zeilingge1988}
A. Zeilinger,  R. G\"a hler, C. G. Shull, W. Treimer, and W.
Hampe,
Rev. Mod. Phys. {\bf 60},1067  (1988). 


\bibitem{Rauch-Werner2000}
H.  Rauch and S.  Werner, {\it Neutron Interferometry: Lessons in Experimental Quantum Mechanics}  (Oxford Univ.  Press, 2000), p. 21;
A. Zeilinger,
Rev. Mod. Phys. {\bf 71}, S288 (1999).

\bibitem{Dirac1928}P. A. M.  Dirac,
{ Proc.  Roy.  Soc. } A {\bf117}, 610 (1928);
  {\bf 118}, 351 (1928).

 \bibitem{Anderson1932}
C. D.  Anderson,  {  Sci. } {\bf 76}, 238 (1932);
{  Phys.  Rev. } {\bf 43}, 491 (1933);
{\bf 43}, 381 (1933).

 \bibitem{Blackett1933}P. M.  S.  Blackett and G.  P.  S.  Occhialini, {  Proc.  Roy.  Soc. } {\bf 139}, 699 (1933).




\bibitem{Thribaud1934}J.  Thibaud, {  Compt.   Rend. } {\bf 197}, 162 (1933); {  Phys.  Rev. } {\bf 45}, 781 (1934).

\bibitem{Joliot1933} F.  Joliot, {\it Compt.  Rend. } {\bf 193}, 1622  (1933); {\it ibid} {\bf 198}, 81 (1934).

\bibitem{Klemperer1934} O.  Klemperer, {  Proc.  Camb.  Phil.  Soc. } {\bf 30},  347 (1934).

\bibitem{Yoshizawa1984}Y.  Yoshizawa {\it et al. },
{  Phys.  Soc.  Jpn. } {\bf 53}, 4125 (1984).



\bibitem{Kaufmann1906}
W.  Kaufmann,
 {  Ann.  der Phys. } {\bf  19}, 495 (1906).

\bibitem{Bucherer1908}
A. H.  Bucherer, {  Phys.  Zeit. } {\bf 9}, 755 (1908);
      {  Ber.  d.  deut.  Phys.  Ges. } {\bf  6}, 688 (1908).
{  Ann.  der Phys. } {\bf 28},  (1909).

\bibitem{Guye:1915}
Ch. E.  Guye and Ch.  Lavanchy, {  Compt.  Rend. } {\bf 161}, 52 (1915).

\bibitem{Elderetal1947} F.  R.  Elder, A.  M.  Gurewitsch, R.  V.  Langmuir and F. D.  Pollack,
{  Phys.  Rev. } {\bf 71}, 829
(1947); {  J.  Appl.  Phys. } {\bf 18}, 810 (1947).





\bibitem{Gillies1997}
G. T.  Gillies,
{  Rep.  Prog.  Phys. } {\bf  60}, 151(1997).







\end{thebibliography}
\end{document}